\documentstyle[12pt]{article}
\begin{document}
\begin{titlepage}
{\hskip 11cm} BIHEP-TH-97-011
\vspace*{2.7cm}
\begin{center}
{\Large Glueball Spectrum from the B.S. Equation\\}
\vspace*{0.4cm}
{{J.Y. Cui $^a$ $^ b$ J.M. Wu $^a$ and H.Y. Jin $^a$}\\
  {$^a$ \small Institute of High Energy Physics}\\
  {\small Beijing, 100039, P.R. China}\\
  {$^b$ \small Department of Physics, Henan Normal University}\\
  {\small Xinxiang, 453002, P.R. China}}
\end{center}
\date{}
\begin{center}
\begin{minipage}{120mm}
\vspace*{1.5cm}
\begin{center}{\bf Abstract}\end{center}
{ The mass of the glueballs is calculated in the B.S. equation framework. 
Under instantaneous 
approximation, the wave function of B.S. equations are obtained. The kernel is 
chosen as the sum of an one-gluon exchange potential, a contact interaction
and a linear confining 
potential. The numerical
results are in agreement with that of 
recent lattice calculation.\\
{\bf PACS: 14.40.C, 11.10.St. }}
\end{minipage}
\end{center}
\vskip 1in
\end{titlepage}
\newpage
1.{\it Introduction}${ ~~} $  The existence of glueball states 
is a very important prediction of QCD, 
whose discovery would be a direct confirmation of the non-abelian
(self-coupling) character of the gluonic degree of freedom in strong
interactions.
Therefore, these states are in focus of the interest of both theorists and
experimentalists. Glueballs were  suggested by Gell-Mann and
Fritzsch\cite{s1}, Fritzsch and Minkowski\cite{s2}. Since then a lot of 
investigation has been made. 
There are various theoretical approaches applied, which include the QCD sum
rules\cite{s3,s4}, bag model\cite{s5,s6}, nonrelativistic potential
model\cite{s7} and lattice calculation\cite{s8,s9}. 
However, these approaches differ from each other markedly in 
their mass prediction,
sometimes as large as 1 GeV. Therefore, it's necessary to make more 
investigations.
\par
On the other hand, in relativistic
quantum field theory, the Bethe-Salpeter(B.S.) equation is an exact
bound state equation. With some appropriate approximation it has been used
successfully in the study of $q\bar q$ bound states\cite{s10}.  
Therefore, it would be reasonable for us to try to study the bound 
states of the gluons in the framework of  B.S. equation.
In our investigation, the B.S kernel is chosen as the sum of an one-gluon 
exchange potential, a contact interaction  and a linear confining potential. 
We calculate the mass of  $0^{++},2^{++},0^{-+}$ and $2^{-+}$ glueballs. 
The numerical results are 
in good agreement with that of recent lattice calculation.
\par
The paper is arranged as follows: in the next section, we construct
the B.S. equation for the bound states of two constituent gluons, and then we
discuss the integral kernel and the structure of the B.S. wave functions 
of glueballs investigated. 
In section three the numerical results and a discussion are presented.
\par
2.{\it The B.S. Equation for Glueballs}${ ~~ }$
  Let $A^a _\mu (x_1)$ and $A^b _\nu (x_2)$ be the gluon fields at
  points $x_1$ and $x_2$,
$|G \rangle $ the bound state of two gluons with mass $M$ and momentum $P_\mu$.
Then the B.S. wave function for a bound state is defined as
\begin{equation}
\label{}
\chi^{a b}_{\mu \nu}(P,x_1,x_2)=\langle 0 |T(A^a _\mu (x_1) A^b
_\nu (x_2))| G \rangle,
\end{equation}
where $\mu ,  \nu $
are Lorentz indices, and $a, b$ color indices. The glueballs are color singlet 
states, so we have
\begin{equation}
\chi^{a b}_{\mu \nu}(P,x_1,x_2)=\delta^{ab}\chi_{\mu \nu}(P,x_1,x_2).
\end{equation}
The translation invariance of system implies
\begin{equation}
\label{}
\chi_{\mu \nu}(P,x_1,x_2)=e^{iP\cdot X} \chi_{\mu \nu}(P,x),
\end{equation}
where $X=\displaystyle\frac{x_1+x_2}{2}$ and $x=x_1-x_2$. We further define 
the B.S. wave function in momentum space as
\begin{equation}
\chi_{\mu \nu}(P,q)=\int \frac{d^4 x}{(2 \pi )^4}e^{-iqx}\chi_{\mu \nu}(P,x),
\end{equation}
where $q$ is the relative momentum of the two constituent gluons.
\par
With a standard method we obtain the B.S. equation for a color singlet
glueball state,
\begin{equation}
\label{e1}
\chi_{\mu \nu}(P,q)=\Delta_{\mu \alpha}(p_1)\Delta_{\nu
\beta}(p_2)\int\frac{d^4 k}{(2\pi)^4}G^{\alpha \beta \rho
\sigma}(P,q,k)\chi_{\rho \sigma}(P,k),
\end{equation}
where the color indices have been suppressed.
 The tensor kernel $G^{\alpha \beta \rho \sigma}(P,q,k)$ in the above equation
is defined as the sum of all two-particle irreducible graphs,
and $\Delta_{\mu \alpha}(p_i)$ is the full propagator of the constituent gluons
with momentum $p_i$, and

\begin{equation}
\label{e2a}
 P=p_1+p_2,
\end{equation}
\begin{equation}
\label{e2b}
 2q=p_1-p_2,
\end{equation}
where $P^2$=$M^2$.
\par
Although this equation is formally exact, it is
difficult to use in concrete problems. The reason is twofold: first
we don't know how to calculate the kernel of B.S. equation, and second, even
if we know the kernel we would be unable to solve the equation exactly. 
Therefore, to make the approximation is inevitable.
\par
In solving the B.S. equation of glueballs, we make ``static'' approximation
as usually do in the case of $q \bar q$ mesons. 
In this approximation one neglects the dependence of the kernel
$G(P,q,k)$ on $k_0$ and $q_0$.
Then the $dk_0$ integration on the right-hand
side of equation (5) yields the equal time wave function
\begin{equation}
\varphi_{\mu \nu}(P,{\bf k})\equiv \int dk_0 \chi_{\mu \nu}(P,k).
\end{equation}
Next we perform $dq_0$ integration on both sides of equation (5), and at the
left-hand side this again yields the three dimensional wave function
$\varphi_{\mu \nu}(P,\bf{q})$. On the right-hand side, the  $q_0$
dependence is only contained in the variables $p_{10}$ and $p_{20}$. That is
\begin{equation}
\varphi_{\mu \nu}(P,{\bf q})=\int dq_0 \Delta_{\mu \alpha}(p_1) 
\Delta_{\nu \beta}(p_2) \int \frac{d^3k}{(2\pi)^4}G^{\alpha \beta \rho
\sigma}(P,{\bf q,k})\varphi_{\rho \sigma}(P,{\bf k}).
\end{equation}
To go on, an appropriate gauge has to be fixed in equation (9).  It's
convenient for us to choose Coulomb gauge. In Coulomb gauge the propagator is
\begin{equation}
\Delta_{\mu \nu}(k)=-\frac{i}{k^2+i\epsilon }\left[g_{\mu
\nu}-\frac{1}{{\bf k}^2}(k_\mu k_\nu - k_\mu s_\nu - k_\nu s_\mu)\right],
\end{equation}
where
\begin{equation}
s_\mu \equiv (0,\bf{k}).
\end{equation}
We can write the propagator more clearly as
\begin{equation}
\Delta_{i j}(k)=\frac{i}{k^2+i\epsilon}(\delta_{i j}-\frac{k_i k_j}{{\bf
k}^2}),
\end{equation}
\begin{equation}
\Delta_{0 0}(k)=\frac{i}{{\bf k}^2},
\end{equation}
\begin{equation}
\Delta_{0 i}(k)=\Delta_{i 0}(k)=0.
\end{equation}
From (9) and (11), one finds that( for $\epsilon_\mu$, the polarization
vector of the constituent gluons, we set $\epsilon_0=0$)
\begin{equation}
\varphi_{0 i}=\varphi_{i 0}=\varphi_{0 0}=0.
\end{equation}
Therefore, in Coulomb gauge one need only deal with the three dimensional
wave functions $\varphi_{i j}(P,{\bf q})(i=1,2,3)$. It's appropriate to give an
effective mass $m$ to the constituent gluons and rewrite the propagator as
\begin{equation}
\Delta_{i j}(k)=\frac{i}{k^2-m^2+i\epsilon}(\delta_{i j}-
\frac{k_i k_j}{{\bf k}^2}).
\end{equation}
For convenience, we adopt the center-of-mass system in the present paper. 
In this frame we have
\begin{equation}
\int \frac{d q_0}{(p_1^2-m^2+i\epsilon)(p_2^2-m^2+i\epsilon)}
=\frac{2 \pi i}{E(4E^2-M^2)},
\end{equation}
where  $E=\sqrt{m^2+{\bf q}^2}$. Substituting (16) and (17) into
equation (9), we obtain
$$
E(M^2-4E^2)\varphi_{ij}(P,{\bf q})=$$
\begin{equation}
i(\delta_{i i^\prime}-\frac{q_i
q_{i^\prime}}{{\bf q}^2})(\delta_{j
j^\prime}-\frac{q_jq_{j^\prime}}{{\bf q}^2})\int \frac{d^3k}{(2\pi)^3}
G_{i^\prime j^\prime k
l}(P,{\bf q,k})\varphi_{k l}(P,{\bf k}).
\end{equation}
This equation is the starting point of our numerical investigation. 
\par
The next question is how to construct the integral kernel. We assume that the
kernel $G_{ijkl}(P,\bf{q,k})$ is the sum of two parts: the short distance part 
$G_{ijkl}^{(s)}(P,\bf{q,k})$, and long distance part $G_{ijkl}^{(l)}(P,\bf{q,k})$. 
As usual,we assume that the short distance part, $G_{ijkl}^{(s)}(P,\bf{q,k})$, 
is contributed by the three lowest order diagrams shown in figure 1.\\  
Besides equations (6) and (7), we also have
\begin{equation}
P=p_3+p_4,
\end{equation}
\begin{equation}
2k=p_3-p_4.
\end{equation}
Calculating diagram $a,b$ and $c$, $G^{(s)}_{\mu \nu \rho \sigma}(P,q,k)$
can be expressed explicitly as
$$
\begin{array}{rl}
\displaystyle G^{(s)}_{\mu \nu \rho \sigma}(P,q,k)&\displaystyle=
3i(4\pi \alpha_s)\left\{2C_{\mu \rho
\tau}(p_1,p_3)C_{\nu \sigma \tau^\prime}(p_2,p_4)
\left[ \frac{g^{\tau 0} g^{\tau^\prime 0}}{{\bf l}^2}
+\frac{g^{\tau i }g^{\tau^\prime j}}{l^2}(\delta_{ij}
-\frac{l_i l_j}{{\bf l}^2}) \right]\right.\\
&\displaystyle\left. -(2g_{\mu \nu}g_{\rho
\displaystyle\sigma}-g_{\mu \rho}g_{\nu \sigma}-g_{\mu \sigma}g_{\nu
\rho})\frac{}{}\right\},
\end{array}
\eqno(21)$$
where $l(=q-k)$ is the momentum exchanged between the two constituent gluons, and
$$
\begin{array}{c}
C_{\mu \rho \tau}(p_1,p_3)=(p_1-2p_3)_\mu g_{\rho \tau}+(p_1+p_3)_\tau
g_{\mu \rho}+(p_3-2p_1)_\rho g_{\mu \tau},
\end{array}
\eqno(22)$$
$$
\begin{array}{c}
C_{\nu \sigma \tau^\prime}(p_2,p_4)=(p_2-2p_4)_\nu g_{\sigma
\tau^\prime}+(p_2+p_4)_{\tau^\prime}
g_{\nu \sigma}+(p_4-2p_2)_\sigma g_{\nu \tau^\prime}.
\end{array}
\eqno(23)$$
The factor 3 on the right-hand side of equation (21) is the color factor which
is $\frac{4}{3}$ in the case of $q\bar q$ bound state.  Diagram $a$ and
diagram $b$ make the same contribution to physical states, so there is
a factor 2 in equation (21). The strong coupling constant $\alpha_s$ is chosen
as a running one,
$$
\begin{array}{c}
\displaystyle\alpha_s=\frac{12\pi}{27}\frac{1}{ln(a+\frac{l^2}{\Lambda^2_{QCD}})},
\end{array}
\eqno(24)$$
where $a$ is a parameter introduced to avoid the infrared divergence. 
As for the long distance part $G^{(l)}_{ijkl}$,
we have no reliable knowledge about it, so
we have to constructed it phenomenally. The experience with $q\bar q$ bound state
shows that the long  distance part and short distance part of kernel have
different spin dependence. Generally speaking, the long distance part makes no
contribution to spin-spin interaction. With the guidance of this experience,
we only chose in equation (22-23) the terms
containing tensor $g_{\mu \rho}g_{\nu \sigma}$ as the spin dependence of the
confining part because such terms have nothing to do with spin effect.
Therefore we assume
$$
\begin{array}{c}
G^{(l)}_{\mu \nu \rho \sigma}=2i(p_1+p_3)\cdot
(p_2+p_4)g_{\mu \rho}g_{\nu \sigma}G(l),
\end{array}
\eqno(25)$$
where $G(l)$ is spatial dependence of confining part of the kernel. We choose
$$
\begin{array}{c}
\displaystyle G(l)=\frac{8\pi\lambda}{l^4},
\end{array}
\eqno(26)$$
which corresponds to a linearly growing potential. The expression
$\frac{1}{l^4}$ is very singular at the zero point of $l$, and
regularization is necessary. The method is to subtract a $\delta$ function
from the confining part of the kernel, that is to make the following
replacement,
$$
\begin{array}{lll}
G^{(l)}_{\mu\nu\rho\sigma} &=& 2i(p_1+p_3)\!\cdot\!(p_2+p_4)g_{\mu\rho}
\displaystyle g_{\nu\sigma}\frac{8\pi\lambda}{(l^2+u^2)^2}\\
&-& \delta^3({\bf l})
\displaystyle\int d^3k\left\{ 2i(p_1+p_3)\!\cdot\!(p_2+p_4)g_{\mu\rho}g_{\nu\sigma}
\frac{8\pi\lambda}{(l^2+u^2)^2} \right\}
\end{array}$$
where $u$ is a small quantity. In actualcalculation we let $u\rightarrow 0$.
In this way the infrared divergence is subtracted out.
\par
The next important question is to construct B.S. wave function for a given
gluon bound state. For $0^{-+}$ glueball the most general decomposition of
the four dimensional wave function  is given by the following expression
$$
\begin{array}{c}
\chi_{\mu \nu}(P,q)=f\epsilon_{\mu \nu \alpha \beta}P^\alpha q^\beta.
\end{array}
\eqno(27)$$
For $0^{++}$ glueball,
$$
\begin{array}{c}
\chi_{\mu \nu}(P,q)=f_{0}g_{\mu \nu}+f_{1}P_\mu P_\nu +f_{2}P_\mu
q_\nu+f_{3}P_\nu q_\mu +f_{4}q_\mu q_\nu,
\end{array}
\eqno(28)$$
where $f$ and $f_{i}$ are scalar functions of $P^2$, $q^2$ and 
$P\cdot q$.  For other states the B.S. wave functions are more complicated
and are not given here. Except for pseudoscalar state, the B.S.
wave function has many independent components. This makes the numerical 
calculation difficult. However, when Coulomb gauge as well as the 
center-of-mass frame is adopted, the wave function becomes simple.
\par
From equation (18), we find
$$
\begin{array}{c}
q_i\varphi_{ij}(P,{\bf q})=0.
\end{array}
\eqno(29)$$
This equation gives a very strong restriction on the wave functions. The
following are some three dimensional wave functions. \\
For $0^{-+}$ state(pseudoscalar)
$$
\begin{array}{c}
\varphi_{ij}(P,{\bf q})=f_p({\bf q})q_k\epsilon_{ijk}.
\end{array}
\eqno(30)$$
For $0^{++}$ state(scalar)
$$
\begin{array}{c}
\varphi_{ij}(P,{\bf q})=f_s({\bf q})(\delta_{ij}-\displaystyle\frac{q_iq_j}{{\bf q}^2}).
\end{array}
\eqno(31)$$
For $2^{++}$ state
$$
\begin{array}{c}
\displaystyle\varphi_{ij}(P,{\bf q})=f_{t1}({\bf q})(\delta_{ik}-\frac{q_iq_k}{{\bf q}^2})
\eta_{kl}(\delta_{jl}-\frac{q_jq_l}{{\bf q}^2})+
f_{t2}({\bf q})(\delta_{ij}-\frac{q_iq_j}{{\bf q}^2})\eta_{kl}q_k q_l.
\end{array}
\eqno(32)$$
For $2^{-+}$ state
$$
\begin{array}{ll}
\varphi_{ij}(P,{\bf q})=&\displaystyle
f_{p1}({\bf q})\epsilon_{klm}\eta_{mn}q_n(\delta_{ik}-\frac{q_iq_k}{{\bf q}^2})
(\delta_{jl}-\frac{q_jq_l}{{\bf q}^2})\\
&\displaystyle+f_{p2}({\bf q})(\epsilon_{kmn}\eta_{ml}q_n-\epsilon_{lmn}\eta_{mk}q_n)
(\delta_{ik}-\frac{q_iq_k}{{\bf q}^2})(\delta_{jl}-\frac{q_jq_l}{{\bf q}^2})\\
&\displaystyle +f_{p3}({\bf q})q_k\epsilon_{ijk}\eta_{mn}q_mq_n,
\end{array}
\eqno(33)$$
where $\eta_{ij}$ is the polarization tensor of the glueball and $f_p$,
$f_s$, $f_{ti}$ and $f_{pi}$ are 
scalar functions of $|{\bf q}|$.
\par
Substituting equation (30) and the expression of kernel (21) and (25) into equation
(18), we obtain an equation about $0^{-+}$ state(For simplicity, in the
following formulae, $|{\bf q}|$ and $|{\bf k}|$ are
expressed simply by $q$ and $k$ respectively.)
$$
\begin{array}{rl}
{~~}&E(M^2-4E^2)f_p(q)=\\
&-8\pi\displaystyle\int \frac{d^3 k}{(2\pi)^3}\left\{\left[A(q,k)
\frac{k}{q}cos\theta
+8k^2sin^2\theta\right]V_s(q,k)
+B(q,k)\frac{k}{q}cos\theta V_l(q,k)\right\}f_p(k),
\end{array}
\eqno(34)$$
where $\theta$ is the angle between vectors ${\bf q}$ and ${\bf k}$, and
$$
\begin{array}{c}
\displaystyle A(q,k)=M^2+q^2+k^2+2qkcos\theta
-\frac{(q^2-k^2)^2}{q^2+k^2-2qkcos\theta },
\end{array}
\eqno(35)
$$
$$
\begin{array}{c}
B(q,k)=M^2+q^2+k^2+2qkcos\theta,
\end{array}
\eqno(36)$$
$$
\begin{array}{c}
\displaystyle V_s(q,k)=\frac{4\pi}{3(q^2+k^2-2qkcos\theta)ln(a+
\frac{q^2+k^2-2qkcos\theta}{\Lambda^2_{QCD}})},
\end{array}
\eqno(37)$$
$$
\begin{array}{c}
\displaystyle V_l(q,k)=\frac{2\lambda}{(q^2+k^2-2qkcos\theta)^2}.
\end{array}
\eqno(38)$$
The equations for other states are given in the Appendix. We can see
that for $0^{-+}$ state and $0^{++}$ state the equations are simple because
each wave function only has one component. However, for $2^{++}$ state we have
to solve two coupled integral equations. The appearance of the wave function
of $2^{-+}$ state is a little awful but the actual calculation shows that
the three coupled equations about it are decoupled into two, 
so it is not too difficult to solve the equations.
\par
3.{\it Results and discussion}${ }$
   There are three parameters in the present model:
$a$ and $\Lambda_{QCD}$ appearing in the running coupling constant
and $\lambda$, the string tension. $\Lambda_{QCD}$ is chosen as $200$ MeV
which is commonly used in various models. The parameter $a$ is chosen as
4.0. Such a choice implies that the  running coupling constant tends to its
largest value 1.0 when the exchanged momentum goes to zero.
We can relate the parameter $a$ to the threshold of the two massive gluon
formation, and such a choice corresponds to a threshold of
$400MeV(2\Lambda_{QCD})$.
As for the string tension, $\lambda$ , it's different from the
case of $q\bar q$ bound states where $\lambda$ is about $0.18$ (GeV)$^2$. As
argued in
ref. \cite{s7}, in the most naive picture, the string tension between two gluons
should be twice that between two quarks because each gluon acts like a $q\bar q$
pair. Therefore, we chose $\lambda=0.36$ (GeV)$^2$. Such a value has also
been used in glueball investigation\cite{s11} and condensation calculation in
QCD sum rules\cite{s12}.
\par
As most of the constituent
models, in our investigation we give an effective mass to the constituent
gluons despite the fact that gluons are massless in QCD Lagrangian. 
We can think that dynamical mass of gluons is 
generated through confinement interaction. One measure of the effective
gluon mass is the energy ($\approx 2m$) necessary to break the string joining
two color-octet sources to materialize a gluon pair. Such a question has
been studied in a lattice calculation\cite{s13} with results
$m\stackrel{>}{\sim}520$ MeV. Non-perturbative continuum studies\cite{s14} also
yield $m=(500\pm 200)$ MeV. In our investigation we give the results when
$m=(0, 400, 600)$ MeV.
\par
With all the parameters determined, we solve the B.S.
equation  numerically. In actual calculation, the momentum integration must
be cut at some value $\Lambda$, and when $\Lambda$ large enough, the results
are independent of it.
The results are given in table 1. 
For comparison, the recent lattice results are also given 
in the table.
\par
We can see that  our results are in good agreement with that of lattice
calculation though there are some uncertainties due to the effective gluon
mass. If we set the gluon mass to zero, the results favor that of the UKQCD
group, while the results with a large gluon mass favor that of the GF11
group.
\par
In the early investigation, a light $0^{-+}$ glueball was preferred. For
example, in MIT-Bag model\cite{s6} $M(0^{-+}$)=0.4 GeV, in potential
model\cite{s7}
$M(0^{-+})=1.4$ GeV, and in QCD sum rules\cite{s4}, $M(0^{-+})=1.7$ GeV. 
However, recent lattice calculation indicates that $0^{-+}$
is a very heavy particle\cite{s8}, $M(0^{-+})=2.3$ GeV. Our investigation 
gives the same result as that on lattice.
\par
In the nonrelativistic potential model. the form of the interaction between
the two constituent gluons is similar to that of present paper, but their
results are much saller. There are two reasons for this fact.Firt, the B.S.
equation is a relativistic equation. More relativistic effects still survive
in our model, even many approximations have been made. Second, the string
tension is an important parameter. In our model, the choice
$\lambda=0.36GeV^2$ is reasonable. 
\par
In fact, the effective mass of the gluon is still unclear. Someone
has argued that the effective gluon mass is the result of confinement
mechanism while the others think it incompatible with
the principle of gauge invariance. 
It's interesting that our numerical results show that the
glueball mass is quite insensitive to the effective mass of the gluons. We
can even chose the gluon mass as zero without much change of the mass of the
glueball.  
\par
It's known that nonrelativistic potential models are very successful in meson
spectrum calculation for heavy quark systems.
One of the reasons of the success is that most mesons
studied can be considered as nonrelativistic systems because the mass of
mesons is quite near the sum of the mass of its constituents. However things
are different for glueballs. Notice the fact, that the glueball mass is much 
larger  than the sum of the effective gluon mass. This means that
glueballs are relativistic systems.  To deal with such systems, perhaps, 
the B.S. equation is a more suitable method.
\par
4.{\it  Acknowledgments.}${~~}$ We wish to thank Prof. Yuan-Ben Dai and Prof.
Xin-Heng Guo for interesting discussions.
\par This work was supported by Chinese National Science Foundation.
\newpage
Appendix\\

\par

The B.S. equation for $0^{++}$ state
$$
\begin{array}{rcl}
E(M^2-4E^2)f_s(q)& =& -8\pi \displaystyle\int
\frac{d^3k}{(2\pi)^3}\left\{\left[\frac{1}{2}A(q,k)(1+cos^2\theta)
 + 4(q^2+k^2)sin^2\theta\right]V_s(q,k)\right .\\
&  & +\displaystyle \left.\frac{1}{2}B(q,k)(1+cos^2\theta)V_l(q,k)
-\frac{3\alpha_s}{2}(3-cos^2\theta)\right\}f_s(k).
\end{array}
\eqno(A.1)
$$
The two coupled B.S. equation for $2^{++}$ state
$$
\begin{array}{rcl}
E(M^2-4E^2)f_{t1}(q)&=&-8\pi \displaystyle\int
\frac{d^3k}{(2\pi)^3}\left\{\left[\frac{1}{4}A(q,k)(1+cos^2\theta)^2
-k^2sin^4\theta\right]V_s(q,k)\right .\\
&&\displaystyle\left .+\frac{1}{4}B(q,k)(1+cos^2\theta)^2 V_l(q,k)
+\frac{3\alpha_s}{4}(1+cos^2\theta)^2 \right\}f_{t1}(k)\\
&&+\displaystyle8\pi \int
\frac{d^3k}{(2\pi)^3}
\left\{\left[\frac{1}{4}A(q,k)k^2sin^4\theta
-2k^4sin^4\theta\right]V_s(q,k)\right.\\
&&\displaystyle\left.+\frac{1}{4}B(q,k)sin^4\theta V_l(q,k)
+\frac{3\alpha_s}{4}k^2sin^4\theta\right\}f_{t2}(k),\\
\end{array}
\eqno(A.2)
$$
$$
\begin{array}{rl}
E(M^2-4E^2)&f_{t2}(q)=8\pi\displaystyle\int
\frac{d^3k}{(2\pi)^3}\left\{\left[\frac{1}{8q^2}A(q,k)(5coc^4\theta-6cos^2\theta
+1)\right .\right .\\
&\displaystyle\left .-\displaystyle\frac{k^2}{2 q^2}(5coc^4\theta-6cos^2\theta+1)-
2(3cos^4\theta -5cos^2\theta +2)\right]V_s(q,k)\\
&\displaystyle\left .+\displaystyle\frac{1}{8q^2}B(q,k)(5coc^4\theta -6cos^2\theta +1)V_l(q,k)
+\frac{3\alpha_s}{8q^2}(5cos^4\theta -18cos^2\theta +5)\right\}f_{t1}(k)\\
&-8\pi\displaystyle \int
\frac{d^3k}{(2\pi)^3}\left\{\left[\frac{k^2}{8q^2}A(q,k)(5cos^4\theta+6cos^2\theta-3)
+2k^2(3cos^2\theta-1)sin^2\theta \right .\right .\\
&\left .+\displaystyle\frac{k^4}{q^2}(5cos^2\theta-1)sin^2\theta\right]V_s(q,k)
+\frac{k^2}{8q^2}B(q,k)(5cos^4\theta+6cos^2\theta-3)V_l(q,k)\\
&\left .+\displaystyle\frac{3\alpha_s k^2}{8q^2}(5cos^4\theta-18cos^2\theta+5)\right\}f_{t2}(k).
\end{array}
\eqno(A.3)
$$
The three coupled  B.S. equation for $2^{-+}$ state
$$
\begin{array}{rl}
E(M^2-4E^2)&(f_{p1}(q)+f_{p2}(q))=-8\pi\displaystyle\int
\frac{d^3k}{(2\pi)^3}\left\{\left[\frac{k}{q}A(q,k)cos\theta
+8k^2sin^2\theta\right]V_s(q,k)\right .\\
&\displaystyle\left.+\frac{k}{q}cos\theta B(q,k)V_l(q,k)\right\}
(cos^2\theta-\frac{1}{2}sin^2\theta)
(f_{p1}(k)+f_{p2}(k))\\
&-8\pi \displaystyle\int \frac{d^3k}{(2\pi)^3}\left\{\left[\frac{k^3}{q}
A(q,k)cos\theta sin^2\theta+
2k^4sin^4\theta\right]V_s(q,k)\right .\\
&\left .+\displaystyle\frac{k^3}{q}A(q,k)cos\theta sin^2\theta V_l(q,k)\right\}f_{p3}(k),
\end{array}
\eqno(A.4)
$$
$$
\begin{array}{rl}
E(M^2-4E^2)&(f_{p1}(q)-2f_{p2}(q))=-8\pi\displaystyle\int
\frac{d^3k}{(2\pi)^3}\left\{\left[\frac{k}{q}A(q,k)cos\theta\right .\right .\\
&\left .\left .\displaystyle+8k^2sin^2\theta\right]V_s(q,k)+
\displaystyle\frac{k}{q}cos\theta B(q,k)V_l(q,k)\right\}(f_{p1}(k)+f_{p2}(k))\\
&-8\pi\displaystyle \int \frac{d^3k}{(2\pi)^3}\left\{\left[\frac{k^3}{q}A(q,k)cos\theta sin^2\theta-
4k^4sin^4\theta\right]V_s(q,k)\right .\\
&\left .+\displaystyle\frac{k^3}{q}A(q,k)cos\theta sin^2\theta
V_l(q,k)\right\}f_{p3}(k),
\end{array}
\eqno(A.5)
$$
$$
\begin{array}{rl}
E(M^2-4E^2)&f_{p3}(q)=
-8\pi\displaystyle\int\frac{d^3k}{(2\pi)^3}\left\{\left[\frac{k^3}{2q^3}A(q,k)
(5cos^3\theta-3cos\theta)\right .\right .\\
&\left .-\displaystyle\frac{2k^4}{q^2}(5cos^4\theta
-6cos^2\theta+1)\right]V_s(q,k)\\
&\left .+\displaystyle\frac{k^3}{2q^3}(5cos^3\theta-3cos\theta)B(q,k)V_l(q,k)\right\}f_{p3}(k).
\end{array}
\eqno(A.6)
$$
where $\theta$ is the angle between vector $\bf{q}$ and $\bf{k }$, and the
functions $A(q,k), B(q,k), V_s(q,k)$ and  $V_l(q,k)$ have been defined in
equation(35--38). 

\newpage
Figuer Caption\\

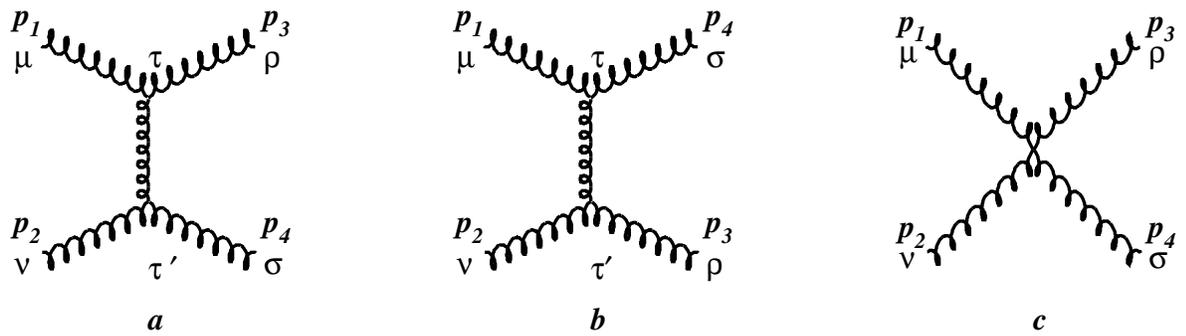
Fig. 1: The diagrams contributing to the short distance part of the B.S.
kernel.

\newpage
Table Caption\\

Table 1: The mass(GeV) of the glueball states. 
\end{document}